# Effects of Nuclear Vibrations on the Energetics of Polythiophene: Quantized Energy Molecular Dynamics

**Sergei Manzhos**

Department of Mechanical Engineering, Faculty of Engineering, National University of Singapore, Blk EA, #07-08, 9 Engineering Drive 1, Singapore 117576

E-Mail: mpemanzh@nus.edu.sg;
Tel.: +65-6516-4605; Fax: +65-6779-1459.

**Abstract:** Effects of nuclear dynamics on the energetics of polythiophene relevant for the performance of organic solar cells are studied for the first time. Nuclear motions change the expectation values of frontier orbital energies and the band gap by about 0.1 eV vs. values at the equilibrium geometry, which is expected to have a significant effect on light absorption, charge separation, and donor regeneration. A new molecular dynamics (MD) algorithm, which accounts for the quantum nature of vibrations, is introduced. It reproduces effects of temperature and deuteration which are lost in the standard MD. Inclusion of quantized vibrations leads to a broadening of the band gap and of energy levels by about 20% at 300K, while having little effect on their expectation values (which change by up to 0.03 eV). Increase in temperature from 300K to 400K and deuteration cause an additional broadening of the spectrum by about 26% and 21%, respectively.

## 1. Introduction

Organic [1] and dye-sensitized solar cells (DSSC) [2] are being actively studied as plausible alternatives to all-semiconductor solar cells, as they may not require high-purity or rare inputs which would inhibit large-scale production. [3,4] In these devices, solar radiation is absorbed by a chromophore molecule – electron donor, which can be purely organic or contain a metal center (usually Ru or Zn). [1,2] The energy levels of the donor are matched to those of the acceptor – either organic (usually C60-based [1]) or semiconductor (usually $TiO_2$ [2]) – in such a way that there is a driving force for electron injection from the excited state of the donor into those of the acceptor. [1,2,5,6] This ensures charge separation. Provided efficient electron / hole collection by negative / positive electrodes, a solar cell is obtained. [1,2] Because the oxidation equivalent hole formed on the donor following the injection needs to be filled, energy level matching between the donor ground state and the positive electrode or a redox species in the electrolyte (in DSSC) [7] is also required to ensure sufficient driving force for donor regeneration.

Key elementary processes governing the operation of organic and dye-sensitized cells – electron injection, charge recombination, chromophore de-excitation - are driven by nuclear motions. [1,5,8-10] Nuclear motions also modulate electronic energy levels, which results in (i) a vibrational broadening of the adsorption spectrum [11] and (ii) changes in energy level matching between the donor and the acceptor. [9,10,12-15] We have shown that nuclear dynamics can cause significant changes in the absorption properties and energy level matching between the donor and the acceptor in DSSC, likely causing changes in the electron injection rate by a factor of 2-3 as well as changes in the recombination rate by orders of magnitude. [12-15] With ab initio molecular dynamics (MD) simulations, it was possible to estimate the distribution over nuclear vibrations of frontier orbitals, the band gap, and driving forces to injection and regeneration. It was found that the expectation values of those distributions may differ from the values computed at the equilibrium geometry by tenths of an eV. Clearly, effects on the solar cell energetics due to nuclear dynamics should be included in theoretical and computational analyses of DSSC and organic cells. To the best of our knowledge, such effects in organic solar cells have not yet been studied.

Molecular dynamics [16-18] simulations, however, provide only a very approximate preview into the effects of nuclear dynamics on the molecular and solar cell energetics. This is largely because they ignore any quantum effects and specifically effects due to the zero-point energy (ZPE). As a result, (i) effects due to isotopic substitution on any coordinate-dependent quantity (e.g. average over vibrations HOLO, LUMO levels) cannot be reproduced as the partition function is mass-independent, and (ii) effects due to temperature are not reproduced for any vibration which does not satisfy $\hbar\omega \ll kT \approx 208\ cm^{-1}$ (at 300K), where $\hbar\omega$ is one quantum of vibrational energy. In fact, this condition is only expected to be satisfied by selected torsional and bending modes and is not satisfied by *all* chemical bonds in practically relevant molecules. Fig. 1 illustrates the severity of the problem; in it, we show the dependence of the vibrational energy (in the harmonic approximation) [19]

$$E_v = \frac{1}{2}\hbar\omega + \frac{\hbar\omega exp\left(-\hbar\omega/kT\right)}{1-exp\left(-\hbar\omega/kT\right)} \qquad (1)$$

on the temperature $T$ for several values of $\hbar\omega$. $\hbar\omega = 3151\ cm^{-1}$ corresponds to a CH bond vibration and $\hbar\omega = 2340\ cm^{-1}$ to a CD bond (these are the frequencies of the CH and CD bond vibrations in the model system considered below). Only at $\hbar\omega = 0$ does the dependence match that of MD, where $E_v = kT$ regardless of the value of $\hbar\omega$. For CH and CD bond vibrations, MD completely ignores that $E_v \gg kT$ due to the ZPE and fails to distinguish between deuterated and non-deuterated bonds. It also imposes an unrealistic temperature dependence of $E_v$. This is important, because $E_v$ and $T$ determine the vibrational amplitude and thereby influence all coordinate-dependent quantities. This situation persists all the way down to $\hbar\omega \approx kT$, i.e. for most bonds. We will show below that MD also fails to maintain an average of $kT$ per degree of freedom in a practical simulation. Quasiclassical trajectory simulations attempt to palliate this by sampling initial conditions from an (approximate) ro-vibrational wavefunction, but do nothing to account for the quantized nature of vibrations along the trajectory. [20] Such sampling is also practically difficult in simulations of large (bio-) molecules. [17]

**Figure 1**. Temperature dependence of the energy stored in vibrational modes (Eq. 1) with energies of 0 ("MD", blue line), 208 cm$^{-1}$ ("kT", 300K, red), 2340 cm$^{-1}$ ("CD", green), and 3151 cm$^{-1}$ ("CH", violet).

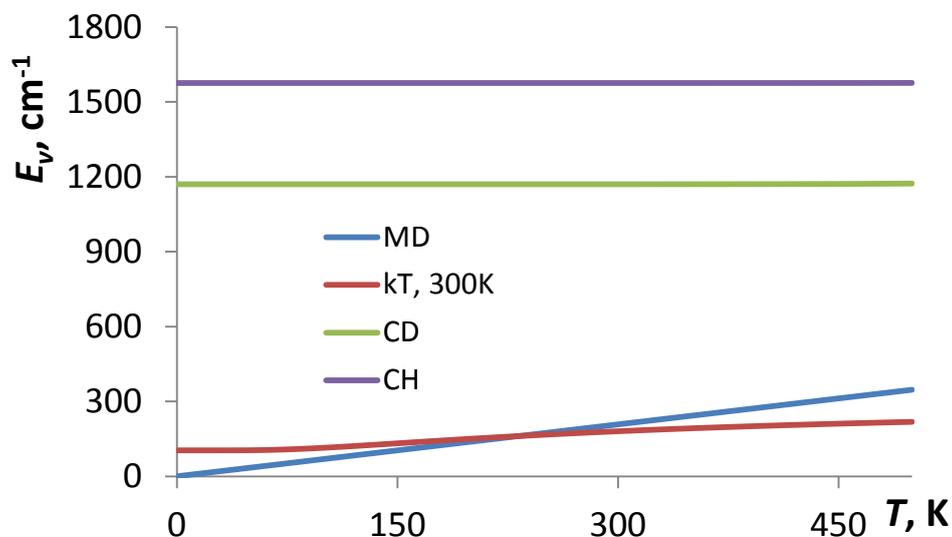

The approximations made in MD are therefore significant in that they limit the descriptive and predictive power of the method. Temperature and isotopic substitution are viable ways to control vibrational and reaction dynamics. [14,15,17,21] Inability of this most widely used method to account for them is therefore a major impediment to rational design of drugs, chemicals, and molecules for organic solar cells, among others.

Here, we propose a method that to a large extent resolves the above problems of MD while maintaining its simplicity. Its basic idea is to maintain a realistic amount of energy in each degree of freedom, in this case, each vibrational mode. We combine it with ab initio MD to study the effects of nuclear motions on the energetics of polythiophene (PT) (Fig. 2) – a widely used building block for donors in organic solar cells [1] – and show that the new method permits observation of effects due to temperature and isotopic substitution which are suppressed in the classic MD approximation. The paper is organized as follows: Section 2 describes the new approach and the details of the ab initio MD setup with which it is coupled, Section 3 presents the results, and Section 4 concludes.

## 2. Theoretical and computational methods

### 2.1. Electronic Structure and Molecular Dynamics Calculations

Electronic and molecular structure was computed using DFT (density functional theory) [22] and the SIESTA code. [23] The PBE exchange-correlation functional [24] and the DZP basis set (double-ζ polarized orbitals) were used. We used a standard DZP basis set as generated by SIESTA, but the cutoff radii were increased from the default values by choosing $E_{shift}$ = 0.01 Ry to mitigate basis-set superposition errors. [25] Core electrons are treated within the effective core approximation with Troullier-Martins pseudopotentials (provided with SIESTA). [26]

Polythiophene (PT) and fully deuterated polythiophene (PT-D) were modeled with a tetramer in a repeated cell (periodic boundary conditions), as shown in Fig. 2. A large enough simulation cell was used that the Brillouin zone was sampled at the Γ point (about 15 x 16 x 15 Å). Geometries were optimized until forces on all atoms were below 0.02 eV/Å. Normal mode frequencies and vectors were computed by diagonalizing the Hessian (dynamic) matrix obtained by finite differences with a step size of 0.03 a.u. The vibrational frequencies of PT and PT-D are shown in Fig. 3. They differ mostly by the vibrations of the CH or CD bonds whose vibrational energies are also shown in Fig. 1 in the classical and quantum harmonic approximations.

**Figure 2**. The structure of polythiophene. The tetramer constituting one simulation cell is shown by balls and sticks and molecules in repeating cells by lines. Atom color scheme: cyan – C, yellow – S, light grey – H/D. Visualization with VMD. [27]

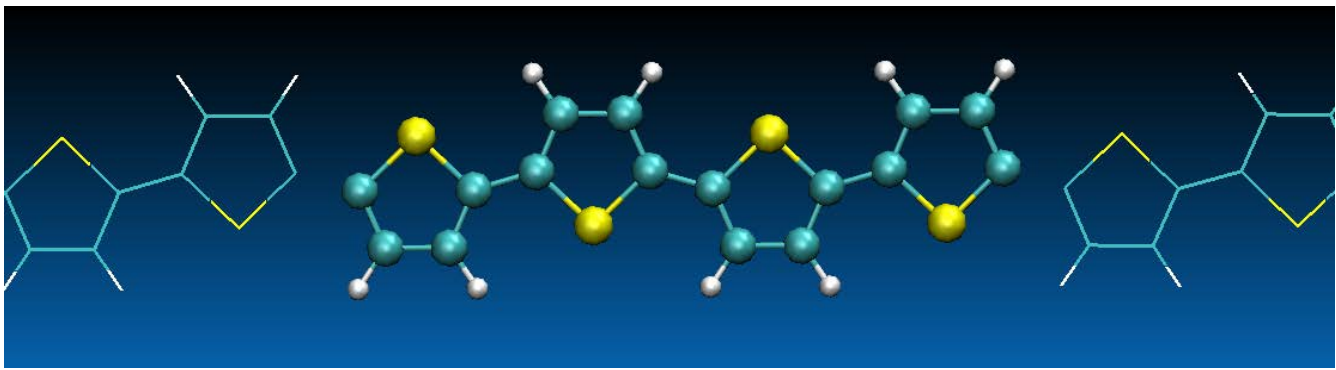

Ab initio molecular dynamics (AIMD) simulations were performed with a time step $dt = 1\,fs$ using velocity rescaling with the relaxation time $\tau = 100\,fs$., i.e. at each time step,

$$\vec{v} = \vec{v}\left(1 + \frac{dt}{\tau}\left(\frac{E_{kin}^{target}}{E_{kin}} - 1\right)\right), \quad (2)$$

where $\vec{v}$ is the vector containing (Cartesian) velocity components of all atoms, $E_{kin}^{target} = \frac{1}{2}N_{DOF}kT$ with $N_{DOF}$ the number of degrees of freedom, and $E_{kin}$ is the kinetic energy of all atoms (excluding system translation). This ensures that the average kinetic energy over a time period of several $\tau$ is $E_{kin}^{target}$ while the system's intramolecular energy redistribution is almost unperturbed (in contrast to instantaneous rescaling). We verified that the results are stable for $\tau$ in the range 50-200 fs. The MD trajectories were 13 ps long, which was sufficient to sample all intramolecular vibrations. We verified that the result do not change when trajectory length changes in the range 10-13 ps. The first $\tau_{eq} = 3$ ps were discarded for equilibration before computing distributions and averages. We verified that the results are stable for $\tau_{eq}$ in the range 2-4 ps. A total of seven dynamical simulations were performed: standard AIMD at 300K and 400K for PT and at 300K for PT-D; QEMD (described in Section 2.2) for PT at 300 K with the average kinetic energy in each mode $\langle E_{kin}\rangle = \frac{1}{2}kT$ (this simulation is directly comparable to MD); QEMD at 300K and 400K for PT and at 300K for PT-D.

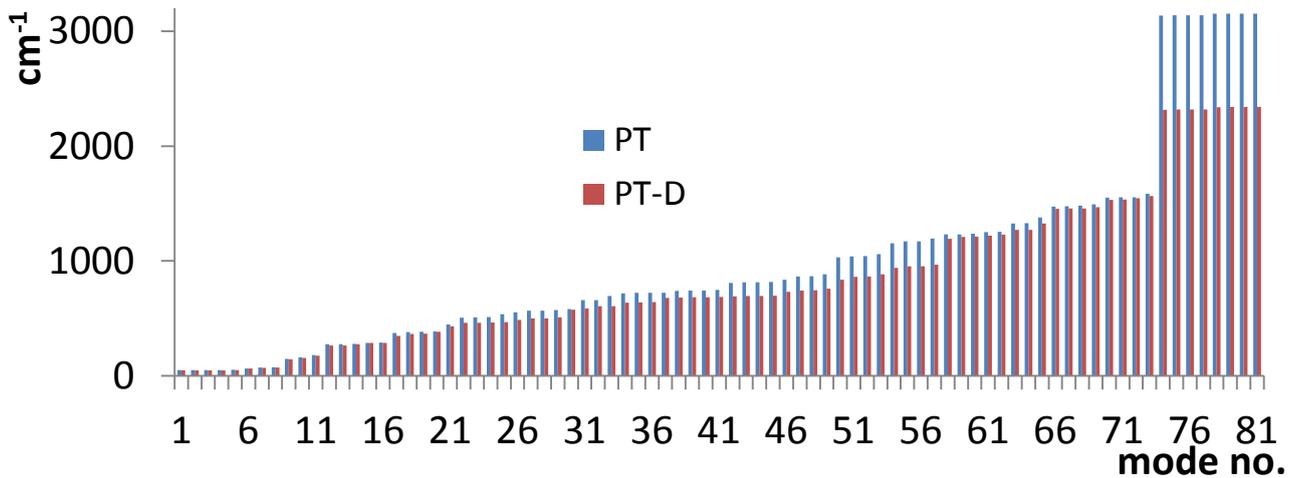

**Figure 3**. The frequencies of the 81 normal modes (excluding translation) of the polythiophene (PT, blue bars) and deuterated polythiophene (PT-D, red) shown in Fig. 2.

*2.2. Quantized Energy Molecular Dynamics*

The problems in MD simulations illustrated by Fig. 1 can be addressed in an approximate way without building the nuclear wavefunction. The basic problem is MD's inability to account for the very different among vibrational modes and very different from $kT$ vibrational energies, due to the ZPE and the quantization of vibrational levels. MD only ensures that the average *total* kinetic energy corresponds to $\frac{1}{2}kT$ per degree of freedom (DOF). It is then *believed* that this also achieves the average kinetic energy of $\frac{1}{2}kT$ in *each* DOF due to efficient redistribution.

Our approach is to make the average kinetic energy in each degree of freedom equal to its ideal, quantum value, i.e. Eq. 1 (hence the name we give to the approach, Quantized Energy MD or QEMD). To this end, we transform from Cartesian to normal mode coordinates obtained in the harmonic analysis as described in Section 2.1:

$$\vec{Q} = \hat{L}^{-1}\hat{M}^{1/2}(\vec{x} - \vec{x}_{eq}), \qquad (3)$$

where $\hat{L}$ is the matrix that diagonalizes the Hessian matrix, $\vec{x}$ is a vector of the Cartesian coordinates of all atoms, $\hat{M}$ is a diagonal matrix containing the masses of all atoms for each component of $\vec{x}$. Similarly, Q-velocities are defined:

$$\vec{v}_Q = \hat{L}^{-1}\hat{M}^{1/2}\vec{v}, \qquad (4)$$

where $\vec{v} = d\vec{x}/dt$. Instead of Eq. 2, we then use

$$v_{Q,i} = v_{Q,i}\left(1 + \frac{dt}{\tau_i}\left(\frac{E_{kin,i}^{target}}{\langle E_{kin,i}\rangle} - 1\right)\right), \qquad (5)$$

where $E_{kin,i} = \frac{1}{2}v_{Q,i}^2$, and $E_{kin,i}^{target} = \frac{1}{2}\left(\frac{1}{2}\hbar\omega_i + \frac{\hbar\omega_i \exp(-\hbar\omega_i/kT)}{1-\exp(-\hbar\omega_i/kT)}\right)$. The brackets stand for time-averaging done at each step according to

$$\langle E_{kin,i}\rangle = \left(1 - \frac{dt}{\tau_i}\right)\langle E_{kin,i}\rangle + \frac{dt}{\tau_i} E_{kin,i} \qquad (6)$$

At the beginning of the simulation, $\langle E_{kin,i}\rangle$ is initialized with $E_{kin,i}^{target}$.

As $\hbar\omega_i$ span many octaves (Fig. 3), we use different relaxation times for different $Q_i$. This is another factor for which the standard MD approach fails to account, as there is typically a single relaxation time, whether velocity rescaling or a Nose thermostat [28] is used. We used

$$\tau_i = \frac{\tau_0}{1+\alpha\frac{\hbar\omega_i - \hbar\omega_{lowest}}{\hbar\omega_{highest}-\hbar\omega_{lowest}}}, \qquad (7)$$

where is $\tau_0 = 100\,fs$. We confirmed that the results reported here are not very sensitive to the choice of $\alpha$ within 10-20. $\alpha = 19$ was used in the calculations reported below. As in the MD simulations, we confirmed that the results are stable for $\tau_0$ in the range 50-200 fs.

After rescaling, the velocities are converted back to the Cartesian space in which forces and displacements are computed. That is, QEMD equations can be inserted into any MD algorithm, whether ab initio or using force fields. We implemented QEMD by modifying the MD part of the SIESTA code.

The implementation of the method as described here uses the harmonic approximation, but it allows for generalization for the anharmonic case. If strong anharmonicity is expected, the expression for $E_v$ can be updated accordingly. It is in principle possible to account for the anharmonic shape of the potential $V(\vec{Q})$ by using a normal mode / HDMR expansion, [29,30]

$$V(\vec{Q}) = \sum_{i=1}^{N_m} V_i(Q_i) + higher\ order\ terms, \qquad (8)$$

where $V_i(Q_i)$ can be anharmonic, and using $E_{v,i} \approx E_{kin}^{anharm}(Q_i) + V_i(Q_i)$ instead of Eq. 1. The method could also be used with other coordinates than normal mode coordinates. Note that most force fields in widely used MD codes represent the potential energy surface (PES) as a sum of uncoupled and often harmonic terms, [31-34] which would be easily amenable to QEMD.

## 3. Results and Discussion

### 3.1. Energetics of Degrees of Freedom and Effective System Temperature

In Fig. 4, distributions of instantaneous temperature (computed from the total kinetic energy of all atoms) are shown for the seven systems. The corresponding averages and standard deviations are listed in Table 1. This is an effective temperature, as opposed to the simulation temperature $T$ in Eq. 1.

**Figure 4**. Distributions of the instantaneous temperature in MD and QEMD runs for PT and PT-D (marked with "–D") at 300K and 400K. QEMD300kT stands for the QEMD run with $\langle E_{kin,i} \rangle = \frac{1}{2}kT$.

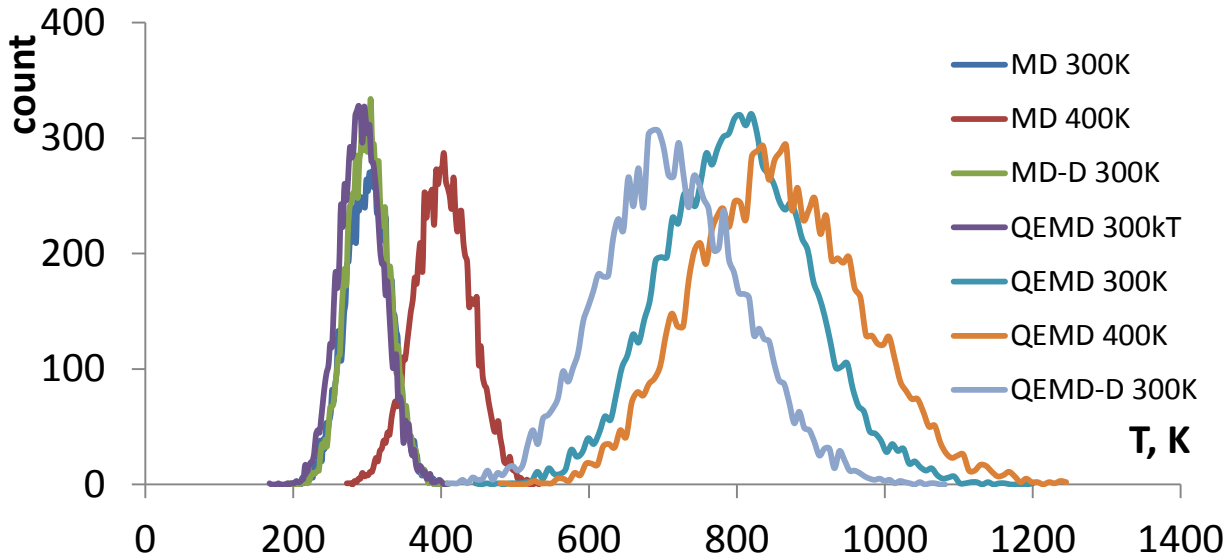

While the average temperatures over the trajectory in the MD simulations are equal to the target temperature to within 0.2 K, the average temperature in the QEMD simulation with $\langle E_{kin,i} \rangle = \frac{1}{2}kT$ in all modes at 300K is about 292 K. The coupling and anharmonicity of the Borh-Oppenheimer PES used by AIMD and the fact that Q-coordinates are non-orthogonal at finite displacements, which leads to "cross-contamination" between modes, may contribute to this discrepancy. Most importantly, however, is that energy redistribution among modes leading to non-uniform and unrealistic (i.e. far from $kT$) mode vibrational energies in MD is prevented by the QEMD algorithm, which also might lead to a different nominal temperature. Figure 5 shows averaged over the trajectories kinetic energies in each DOF, both in Cartesian and in Q-coordinates, obtained from MD and QEMD with $E_{kin,i}^{target} = \frac{1}{2}kT$ in all modes (except translational modes, for which $E_{kin,i}^{target} = 0$ is used). Clearly, QEMD does a much better job than MD at maintaining the average kinetic energy per mode at $\frac{1}{2}kT$: the standard deviation of mode energies (excluding translational modes) in MD is about 31% of the mean ($\frac{1}{2}kT$), while in QEMD, it is only 4%. In the Cartesian coordinates, MD results in the standard deviation of about 18% of the mean, but even here, QEMD is slightly better with the standard deviation of about 15% of the mean. QEMD also keeps mode energies for translation at much lower values (of the order of 0.01 cm$^{-1}$) than MD (0.5-6.5 cm$^{-1}$).

**Table 1**. Averages $\langle T \rangle$ and standard deviations STD, in K, of the instantaneous temperature for all trajectories. Trajectory labeling is the same as in Fig. 4.

|                    | MD300K | MD400K | MD-D300K | QEMD300kT | QEMD300K | QEMD400K | QEMD-D300K |
|--------------------|--------|--------|----------|-----------|----------|----------|------------|
| $\langle T \rangle$ | 300    | 400    | 300      | 292       | 795      | 849      | 710        |
| STD                | 29     | 38     | 28       | 30        | 98       | 111      | 95         |

Maintaining mode energies at $E_v$ leads to higher average kinetic energies and effective temperatures. This is expected given that $E_v(T) > kT$ for modes with $\hbar\omega > kT$, see Fig. 1, and that this condition is true for most modes at the temperatures considered here (Fig. 3). Interestingly, it has been known that MD may have to be done at a higher temperature than $T$ to reproduce structures observed at $T$.[35] It also follows from Fig. 1 that the effective temperature should change less with $T$ in QEMD compared to MD, as for modes with $\hbar\omega \gg kT$, $E_v$ approaches $\frac{1}{2}\hbar\omega$ and is almost independent of $T$. This is indeed observed in Fig. 4 and Table 1: the relative change in effective temperature when $T$ is increased from 300 to 400 K is smaller in QEMD than in MD.

**Figure 5**. Averaged over trajectories energies of the vibrational modes (top) and of the Cartesian degrees of freedom (bottom) for $T$=300 K. Blue empty triangles: MD, filled red circles: QEMD with $E_{kin,i}^{target} = \frac{1}{2}kT$ in all modes. The connecting lines are to guide the eye.

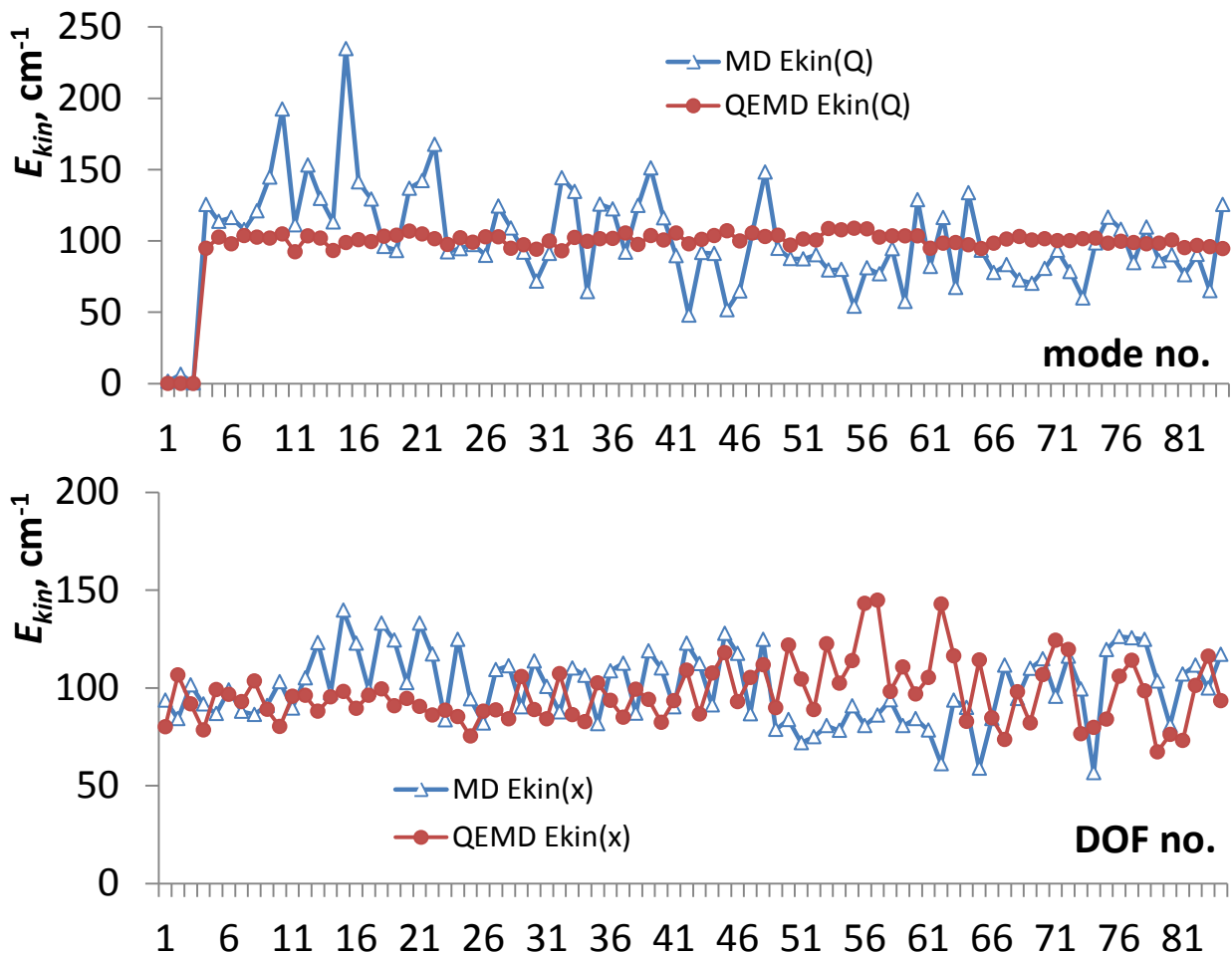

As expected, in MD, there is no effect of the deuteration. In QEMD, deuteration has a strong effect on the average kinetic energy. The vibrational amplitudes of modes with $\hbar\omega > kT$ should be smaller in the deuterated molecule, which in the classical approximation should translate into a lower average kinetic energy and consequently lower effective temperature. This is exactly what is observed in QEMD.

**Figure 6**. Lines without symbols: target mode average kinetic energies corresponding to vibrational energies of Eq. 1 for PT at 300K (solid black line, "target-H 300K") and 400K (dashed black line, "target-H 400K") and for PT-D at 300K (dashed gray line, "target-D 300K"). Symbols: average mode kinetic energies achieved in QEMD simulations for PT at 300K (open blue triangles, "QEMD 300K") and 400K (filled red circles, "QEMD 400K") and for PT-D at 300K (open green rhombs, "QEMD-D 300K"). The connecting lines are to guide the eye.

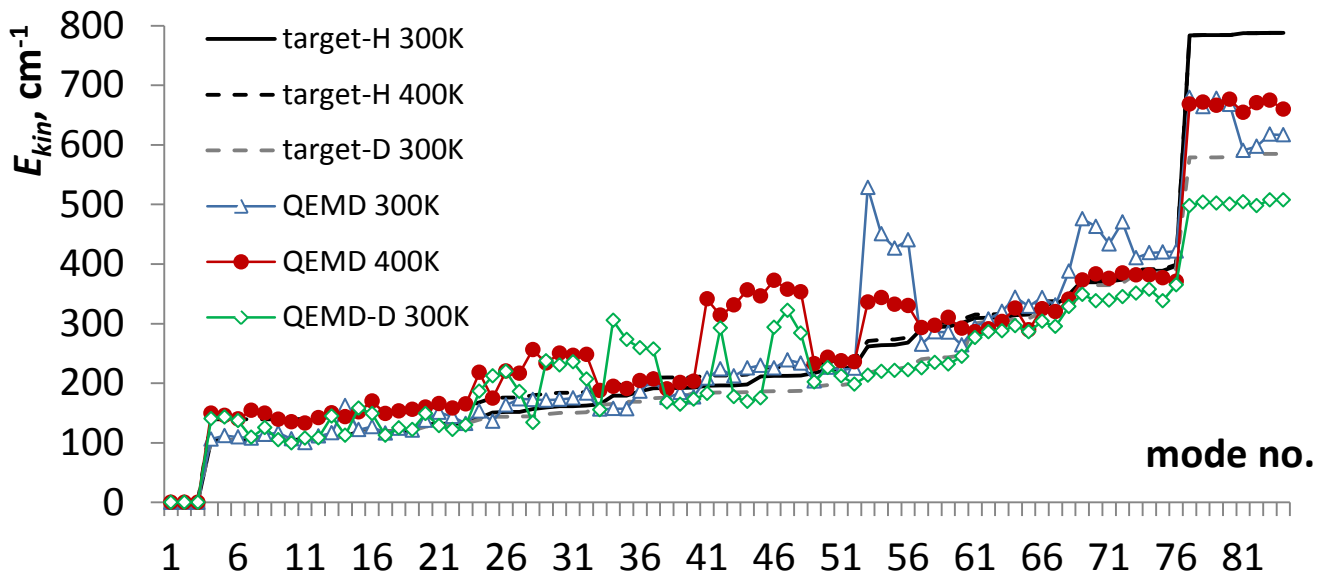

The averaged over the trajectories mode kinetic energies when using Eq. 5 are shown in Fig. 6 for both PT (for $T$ =300K and 400K) and PT-D (for $T$=300K) together with their respective $E_{kin,i}^{target}$. One can appreciate from Fig. 6 how different the mode energies are among themselves and from $\frac{1}{2}kT$, cf. Fig. 5. This difference is completely ignored in the standard MD approach. Of note is the difference between PT and PT-D, also ignored in MD. QEMD simulations generally reproduce the trends in $E_{kin,i}^{target}$, although for some modes, there are significant deviations; specifically, for the highest energy modes (corresponding to the oscillations of the CH and CD bonds), the algorithm does not quite reach the required target energy. This is not surprising given that the algorithm attempts to counter significant redistribution of energy from these modes at time scales comparable to $\tau_i$. These deviations are, however, not larger relative to the target energy than the deviations observed in standard MD, see Fig. 5. The algorithm recovers most of the mode average kinetic energy increase relative to $\frac{1}{2}kT$. For example, the ratio of the energies in the highest 8 modes corresponding to CH/CD stretches in QEMD simulations of PT and PT-D is 1.27 on the average vs. 1.35 for the ratio of $E_{kin,i}^{target}$, meaning that with QEMD, it is possible to simulate effects of isotopic substitution. QEMD also recovers the correct temperature dependence of mode energies following from Eq. 1 and Fig. 1: an increase of $T$ from 300K to 400K has very little effect on the highest-energy modes, while for the (six) modes with $\hbar\omega < 100\ cm^{-1}$, the ratio of $\langle E_{kin,i}\rangle$ is 1.31 – mimicking the ratio of the temperatures, as expected.

10Finally, the standard deviation of the instantaneous temperature is about 10% of the mean in MD and in QEMD with $\langle E_{kin,i}\rangle = \frac{1}{2}kT$ in all modes; it is about 12-13% in QEMD using Eq. 5. This is not surprising given the larger range of atomic and mode kinetic energies when using $E_v$.

*3.2. Effect of Nuclear Vibrations on the Absorption Spectrum and Energy Levels of Polythiophene*

The higher $\langle E_{kin,i}\rangle$ in modes with $\hbar\omega > kT$ realized in QEMD as described in the previous section means that that geometry distortions are larger than in MD due to larger oscillation amplitudes of high frequency modes. This is because $E_{kin,i}^{max} = E_{potential,i}^{max} = \frac{1}{2}Q_{i,max}^2$, in the harmonic approximation. These distortions from the equilibrium geometry are large contributors to the vibrational broadening of the absorption spectrum. [11,36] In the present system, the first peak in the visible absorption spectrum is due to the HOMO→LUMO transition. The frontier orbitals are shown in Fig. 7 We have shown previously that when the absorption peak of organic dyes is due to the HOMO→LUMO transition, its shape can be modeled as the distribution of the HOMO-LUMO gap resulting from nuclear vibrations. [11,36] Further, the distributions of the energies of LUMO and HOMO levels themselves are important, as they determine electron injection and donor regeneration, respectively, in solar cells using organic donors. [9,10,12-15]

**Figure 7**. HOMO (left) and LUMO (right) orbitals of the polythiophene. The absolute values are plotted. The lines denote the edges of the periodic simulation cell.

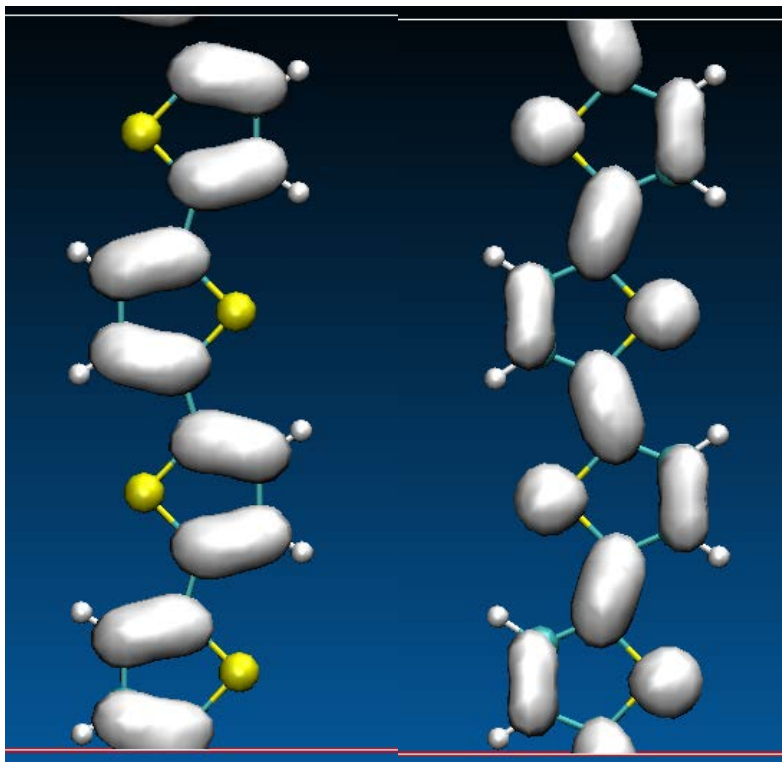

In Table 2, we list averaged over the trajectories values of HOMO, LUMO energy levels and of the HOMO-LUMO gap for the seven simulations. For comparison, the values at the equilibrium geometry (which are the same for all systems) are also given. The distributions of the gap are also shown in Fig.

8. Standard deviations are also given in Table 2 as a measure of the width of the absorption spectrum and of the energy levels.

Table 2. Values of the frontier orbital (HOMO, LUMO) energy levels and of the HOMO-LUMO gap, in eV, for all studies systems as well as at the equilibrium geometry. Their standard deviations (where applicable) are given in parentheses.

|      | Equil. | MD 300K | MD 400K | MD-D 300K | QEMD 300kT | QEMD 300K | QEMD 400K | QEMD-D 300K |
|------|--------|---------|---------|-----------|------------|-----------|-----------|-------------|
| HOMO | -4.07  | -4.17(0.06) | -4.18(0.09) | -4.16(0.09) | -4.15(0.09) | -4.17(0.11) | -4.20(0.14) | -4.17(0.12) |
| LUMO | -2.90  | -2.89(0.06) | -2.90(0.09) | -2.89(0.10) | -2.88(0.08) | -2.93(0.10) | -2.92(0.13) | -2.93(0.13) |
| gap  | 1.17   | 1.28(0.12) | 1.28(0.16) | 1.28(0.18) | 1.27(0.16) | 1.25(0.19) | 1.28(0.24) | 1.24(0.23) |

The effects observed in Table 2 and Fig. 8 are: (i) the effect of nuclear motions on the expectation values of the levels and the gap, (ii) the effect of the temperature, (iii) the effect of isotopic substitution, and (iv) the effect of using the quantum $E_v$, i.e. QEMD vs. MD, and how it affects (i-iii). First, nuclear motions change significantly, by up to 0.1 eV, the expectation values of HOMO and of the gap, in all systems. This is expected to have a strong effect on light absorption properties and regeneration. The increase of more than 9% of the expectation value of the gap vs. the equilibrium value would translate into a blue-shift of the order of 50 nm when taking into account the well-known underestimation of the gap by the PBE functional. [11,37] The decrease of the expectation value of HOMO could double the rate of regeneration. [7,38] The expectation value of the LUMO is little changed from its equilibrium value, implying little change in the average driving force to injection vs. the equilibrium value. However, the width of the distribution of LUMO energies of up to 0.13 eV implies that electron injection can still be impacted due to the non-linear effect of the driving force on the injection rate. [1,6] These findings are in line with previous calculations that predicted similar effects of nuclear motions in dye-semiconductor systems. [12-15]

In the standard MD simulation, a temperature increase from 300K to 400K and deuteration have both the effect of increasing the width of the distributions but not changing the expectation values. The changes for the expectation values of the order of 0.01 eV are within the accuracy of the simulations. This is again similar to the MD results reported previously. [14,15] In fact, the absence of the isotopic substitution effect on the expectation values is imposed by the mass-independence of the partition function in MD, and, therefore, any differences between the columns "MD 300K" and "MD-D 300K" in Table 2 can be used to estimate the lower bound on the accuracy. The QEMD simulation for PT at 300K with $E_{kin,i}^{target} = \frac{1}{2}kT$ in all modes reproduces the MD results to within 0.01 eV for LUMO and the gap, only the expectation value of the HOMO level differs by 0.02 eV.

The imposition of a realistic $E_v$ in each mode (Eq. 1) has the consequence of broadening the absorption spectrum by about 20% (Table 2). The spectrum is further broadened by about 26% when the temperature is increased from 300K to 400K. This also leads to the broadening of the distributions of the HOMO and LUMO energies. There is a noticeable drift lower of the expectation value of the LUMO at both temperatures vs. MD, and a lowering of that of the HOMO at 400K. These explain a red-shift of the gap of PT by about 0.03 eV at 300K vs. MD. This is a change of about 2.5% which would result in the shift of the absorption peak maximum by about 13 nm. This is a small effect. The small effect of $E_v$ on the expectation value of the gap is explained by the fact that HOMO and LUMO are centered on the carbon backbone rather than on CH/CD bonds (Figure 7). This is also corroborated by the fact that a higher temperature makes the dynamics effect of the gap stronger, at it results in stronger distortions in the lower-frequency backbone vibrational modes which are sensitive to $T$, while CH/CD modes are not (see Figs. 1 and 6).

**Figure 8**. The distributions of the HOMO-LUMO gap following from MD (top) and QEMD (bottom) simulations. "QEMD 300kT" is for the QEMD simulation with $E_{kin,i}^{target} = \frac{1}{2}kT$ in all modes.

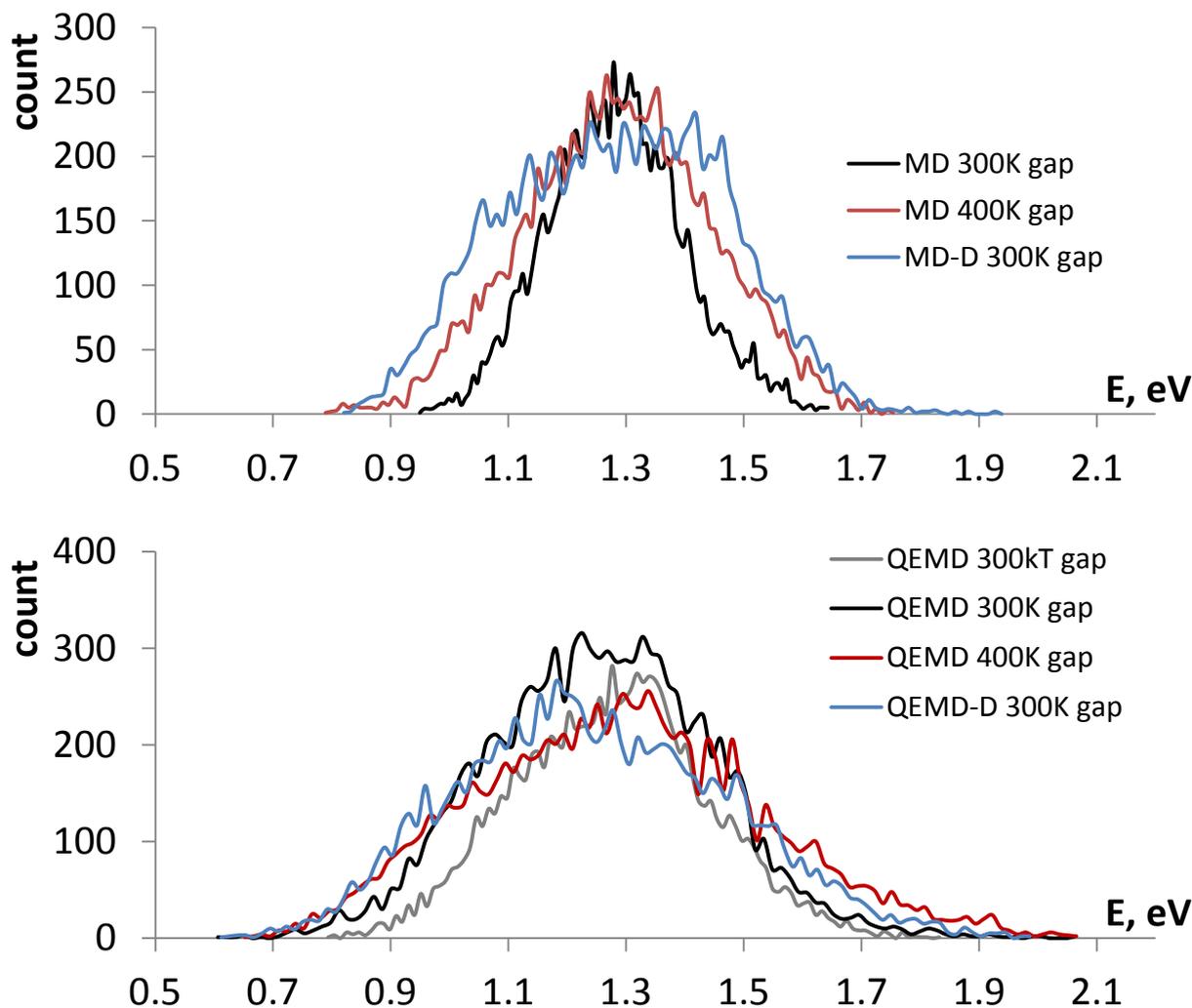

Deuteration has little effect on the expectation values in QEMD but also leads to a larger broadening of both the absorption spectrum (by about 21%) and the distributions of HOMO, LUMO energies (Table 2). This in our view is due a similar mechanism as the temperature effect, specifically, heavier D atoms cause larger displacements of C atoms (as is easily verified by looking at the components of the matrix $\hat{L}$) but do not change their average positions.

## 4. Conclusions

We have presented a modified molecular dynamics algorithm, QEMD (Quantized Energy Molecular Dynamics), which largely fixes the inability of the standard molecular dynamics (MD) approach to account for the quantized nature of vibrations, which in turn leads to inability to model effects of isotopic substitution and of temperature changes. In QEMD, the vibrational energy of each normal mode of vibration $E_v$ is kept at its quantum value, Eq. 1. As a result, simulated mass- and temperature- dependences are realistic.

We have applied QEMD to polythiophene (PT), which is a common building block for many polymer donors in organic solar cells (e.g. P3HT). [1] To test the algorithm, we first performed simulations with target average kinetic energies $\langle E_{kin} \rangle = \frac{1}{2}kT$ in each mode, similar to MD. We have confirmed that QEMD maintains $\langle E_{kin} \rangle$ at $\frac{1}{2}kT$ per mode with a much better accuracy than MD. We then applied QEMD with $E_v$ of Eq. 1 to study the effects of deuteration and temperature on the energetics of PT. We confirmed that QEMD simulations reproduce the dependence of $E_v$ on mode energy for vibrational frequencies spanning many octaves. The relative energetics of protonated and deuterated PT and its temperature dependence are fully reproduced.

Accounting for the real vibrational energy leads to a vibrational broadening of the absorption spectrum (proxied as the band gap) by about 20% at 300K. An increase in temperature to 400K leads to an additional broadening by about 26%. Deuteration also has the effect of broadening the spectrum by about 21% at 300K. Energy levels of frontier orbitals which are critical for energy level matching with the acceptor and electrolyte species are similarly broadened. Nuclear dynamics leads to changes in the expectation values of the band gap and of the HUMO by about 0.1 eV vs. their values at the equilibrium geometry, which is expected to affect significantly the overlap with the solar spectrum and donor regeneration. The computed broadening of the levels by up to 0.14 eV is expected to affect light absorption, electron injection, and regeneration. This is the first time effects of nuclear vibrations on the energetic and energy level matching were studied for an organic solar cell component.

Inability to account for quantum nature of vibrations significantly limits explanatory and predictive power of MD and inhibits molecular design for a wide range of applications, including organic photovoltaics, drug design, and synthesis. As QEMD effectively addresses this problem, we hope it will become useful in all these applications and empower rational molecular and material design.

**Acknowledgments**

This work was supported by the Tier 1 AcRF grant from the Ministry of Education of Singapore.

**References**

[1] T. M. Clarke, J. R. Durrant, *Chem. Rev.* **2010**, *110*, 6736.
[2] A. Hagfeldt, G. Boschloo, L. Sun, L. Kloo, H. Pettersson, *Chem. Rev.* **2010**, *110*, 6595.
[3] C. S. Tao, J. Jiang, M. Tao, *Solar Energy Mater. Solar Cells* **2011**, *95*, 3176.
[4] P. K. Nayak, G. Garcia-Belmonte, A. Kahn, J. Bisquert, D. Cahen, *Energy Environ. Sci.* **2012**, *5*, 6022.
[5] N. Martsinovich, A. Troisi, *Energy Environ. Sci.* **2011**, *4*, 4473.
[6] A. Listori, B. O'Regan, J. R. Durrant, *Chem. Mater.* **2011**, *23*, 3381.

[7] M. Wang, C. Graetzel, S. M. Zakeeruddin, M. Graetzel, *Energy Environ. Sci.* **2012**, *5*, 9394.

[8] V. Coropceanu, J. Cornil, D. A. da Silva Filho, Y. Olivier, R. Silbey, J.-L. Bredas, *Chem. Rev.* **2007**, *107*, 926.

[9] W. R. Duncan, O. V. Prezhdo, *Annu. Rev. Phys. Chem.* **2007**, *58*, 143.

[10] O. V. Prezhdo, W. R. Duncan, V. V. Prezhdo, *Progr. Surf. Sci.* **2009**, *84*, 30.

[11] S. Manzhos, H. Segawa, K. Yamashita, *Chem. Phys. Lett*. **2012**, *527*, 51.

[12] S. Manzhos, H. Segawa, K. Yamashita, *Phys. Chem. Chem. Phys.* **2012**, *14*, 1749.

[13] S. Manzhos, H. Segawa, K. Yamashita, *Proc. SPIE* **2012**, *8438*, 843814.

[14] S. Manzhos, H. Segawa, K. Yamashita, *Phys. Chem. Chem. Phys.* **2013**, *15*, 1141.

[15] S. Manzhos, H. Segawa, K. Yamashita, *Computation* **2013**, in print.

[16] D. Marx, J. Hutter, *Ab Initio Molecular Dynamics: Basic Theory and Advanced Methods* **2009** (Cambridge University Press: Cambridge, UK).

[17] S. A. Adcock, J. A. McCammon, *Chem. Rev.* **2006**, *106*, 1589.

[18] D. C. Rapaport, *The Art of Molecular Dynamics Simulation* **2004** (Cambridge University Press: Cambridge, UK).

[19] D. A. McQuarrie, J. D. Simon, *Physical Chemistry: A Molecular Approach* **1997** (University Science Books: Herndon, VA).

[20] G. D. Billing, *Dynamics of Molecule Surface Interactions* **2000** (John Wiley & Sons: New York NY).

[21] S. Manzhos, J. Fujisawa, H. Segawa, K. Yamashita, *Jpn. J. Appl. Phys.* **2012**, *51*, 10NE03.

[22] W. Kohn, L. J. Sham, *Phys. Rev.* **1965**, *140*, A1133.

[23] J. M. Soler, E. Artacho, J. D. Dale, A. Garcia, J. Junquera, P. Ordejon, D. Sanchez-Portal, *J. Phys.: Condens. Matter.* **2002**, *14*, 2745.

[24] J. P. Perdew, K. Burke, M. Ernzerhoff, *Phys. Rev. Lett*. **1996**, *77*, 3865.

[25] E. Artacho, E. Anglada, O. Dieguez, J. D. Gale, A. Garcia, J. Junquera, R. M. Martin, P. Ordejon, J. M. Pruneda, D. Sanchez-Portal, J. M. Soler, *J. Phys.: Condens. Matter* **2008**, *20*, 064208.

[26] N. Troullier, J. L. Martins, *Phys. Rev. B* **1991**, *43*, 1993.

[27] W. Humphrey, A. Dalke, K. Schulten, *J. Molec. Graphics* **1996**, *14*, 33.

[28] S. Nose, *J. Chem. Phys*. **1984**, *81*, 511.

[29] S. Carter, A. R. Sharma, J. M. Bowman, *J. Chem. Phys.* **2013**, *137*, 154301.

[30] S. Manzhos, T. Carrington, *J. Chem. Phys*. **2006**, *125*, 084109.

[31] W. D. Cornell, P. Cieplak, C. I. Bayly, I. R. Gould, K. M. Merz, D. M. Ferguson, D. C. Spellmeyer, T. Fox, J. W. Caldwell, P. A. Kollman, *J. Am. Chem. Soc.* **1995**, *117,* 5179.

[32] D. A. Case et al., *AMBER 9* **2006** (University of California: San Francisco CA).

[33] B. R. Brooks, C. L. Brooks III, A. D. Mackerell, L. Nilsson, R. J. Petrella, B. Roux, Y. Won, G. Archontis, C. Bartels, S. Boresch, A. Caflisch, L. Caves, Q. Cui, A. R. Dinner, M. Feig, S. Fischer, J. Gao, M. Hodoscek, W. Im, K. Kuczera, T. Lazaridis, J. Ma, V. Ovchinnikov, E. Paci, R. W. Pastor, C. B. Post, J. Z. Pu, M. Schaefer, B. Tidor, R. M. Venable, H. L. Woodcock, X. Wu, W. Yang, D. M. York, M. Karplus, *J. Comp. Chem.* **2009**, *30*, 1545.

[34] J. Wang, R. M. Wolf, J. W. Caldwell, P. A. Kollman, D. A. Case, *J. Comput. Chem*. **2004**, *25*, 1157.

[35] S. Manzhos, *MRS Commun*. **2013**, *3*, in print, DOI: http://dx.doi.org/10.1557/mrc.2012.34

[36] S. B. Rempe, T. R. Mattsson, K. Leung, *Phys. Chem. Chem. Phys.* **2008**, *10*, 4685.
[37] H. Xiao, J. Tahir-Kheli, W. A. Goddard, III, *J. Phys. Chem. Lett.* **2011**, *2*, 212.
[38] T. Daeneke, A. J. Mozer, Y. Uemura, S. Makuta, M. Fekete, Y. Tachibana, N. Koumura, U. Bach, L. Spicca, *J. Am. Chem. Soc.* **2012**, *134*, 16925.